# Iso-oriented monolayer α-MoO$_3$(010) films epitaxially grown on SrTiO$_3$(001)

Yingge Du,[a] Guoqiang Li,[b,c] Erik W. Peterson,[d] Jing Zhou,[d] Xin Zhang,[b] Rentao Mu,[b] Zdenek Dohnálek,[b] Mark Bowden,[a] Igor Lyubinetsky,[a] and Scott A. Chambers[b]

[a]. *Environmental Molecular Sciences Laboratory, Pacific Northwest National Laboratory, Richland, WA, 99352 USA  Email: Yingge.Du@pnnl.gov*
[b]. *Physical and Computational Sciences Directorate, Pacific Northwest National Laboratory, Richland, WA, 99352 USA*
[c]. *Key Laboratory of Photovoltaic Materials of Henan Province, School of Physics & Electronics, Henan University, Kaifeng, 475004 P.R. China*
[d]. *Department of Chemistry, University of Wyoming, Laramie, WY 82072 USA*

**Abstract:**

The ability to synthesis well-ordered two-dimensional materials under ultra-high vacuum and directly characterize them by other techniques *in-situ* can greatly advance our current understanding on their physical and chemical properties. In this paper, we demonstrate that iso-oriented α-MoO$_3$ films with as low as single monolayer thickness can be reproducibly grown on SrTiO$_3$(001) (STO) substrates by molecular beam epitaxy ( (010)$_{MoO3}$ ∥ (001)$_{STO}$, [100]$_{MoO3}$ ∥ [100]$_{STO}$ or [010]$_{STO}$) through a self-limiting process. While one in-plane lattice parameter of the MoO$_3$ is very close to that of the SrTiO$_3$ ($a_{MoO3}$ = 3.96 Å, $a_{STO}$ = 3.905 Å), the lattice mismatch along other direction is large (~5%, $c_{MoO3}$ = 3.70 Å), which leads to relaxation as clearly observed from the splitting of streaks in reflection high-energy electron diffraction (RHEED) patterns. A narrow range in the growth temperature is found to be optimal for the growth of monolayer α-MoO$_3$ films. Increasing deposition time will not lead to further increase in thickness, which is explained by a balance between deposition and thermal desorption due to the weak van der Waals force between α-MoO$_3$ layers. Lowering growth temperature after the initial iso-oriented α-MoO$_3$ monolayer leads to thicker α-MoO$_3$(010) films with excellent crystallinity.

**Introduction**

Layered nanomaterials, such as graphene, $Bi_2Se_3$, $MoS_2$ and α-$MoO_3$, are of significant interest due to their intriguing physical properties and diverse range of applications based on their two dimensional (2D) character.[1-8] In α-$MoO_3$ (space group $P_{bnm}$; lattice constants a = 3.70 Å, b = 13.86 Å, and c = 3.96 Å (JCPD file: 05-0508), edge and corner sharing $MoO_6$ octahedra are linked to create layered sheets that are stacked in the [010] direction, as shown in figure 1a. Lamellar formation is made by linking the adjacent layers along the (010) plane through weak van der Waals forces, while the internal interactions between atoms within each layer are dominated by strong covalent and ionic bonding.[6] Because of its 2D layered structure, α-$MoO_3$ is capable of reversibly absorbing large quantities of foreign atoms, such as Li or Na, within the van der Waals gaps, making it an ideal material for use in electrochromic devices and lithium ion batteries.[9-15] In addition, α-$MoO_3$ has demonstrated potential to be useful in thin-film pseudocapacitors,[16] light-emitting diodes,[17] plasmon resonance generators,[18] and biosensors.[19]

Nanostructures and thin films of α-$MoO_3$ have been synthesized in a variety of ways including hydrothermal, vapor-transport, oxidation of $MoS_2$ nanosheets, and physical evaporation.[15, 20-24] Molecular beam epitaxy (MBE), offering accurate control over the purity, orientation, and thickness, is in principle an ideal way to synthesize ultra-thin, highly oriented $MoO_3$ layers. Additionally, MBE is readily combined in ultra-high vacuum (UHV) with various *in-situ* spectroscopy and microscopy techniques, such as reflection high-energy electron diffraction (RHEED), X-ray photoelectron spectroscopy (XPS), and scanning tunneling microscopy (STM).[25, 26] These methods yield important information critical to accurate understanding of the physical and chemical properties. However, preparation of phase-pure, ultra-thin α-MoO$_3$ films by MBE has proven to be difficult because besides the thermodynamically stable orthorhombic α-phase, a metastable monoclinic β-phase can also crystallize.[6, 9] MoO$_3$ film properties have been shown to be very sensitive to growth temperature.[23, 27-29] Another difficulty lies in the large difference between the two in-plane lattice parameters ($a$ = 3.96 Å and $c$ = 3.70 Å) in α-MoO$_3$ which makes it challenging to identify lattice matched substrates. Commercially available oxide substrates with perovskite (ABO$_3$) structures offer a range of selection with pseudo-tetragonal or pseudo-cubic $a$-axis parameters ranging from 3.7 Å to 4.0 Å,[30] among which SrTiO$_3$ (STO, $a$ = 3.905 Å) provides good match along one in-plane direction. Confirmation of successful α-MoO$_3$ heteroepitaxy on a substrate of cubic symmetry from RHEED has not been reported in the literature. Due to the large difference between $a$ and $c$, if the film nucleates as a single domain and is coherent with the substrate along [100] (i.e. [100]$_{MoO3}$ ∥ [100]$_{STO}$), a different streak spacing is expected along [010], corresponding to partial or complete film relaxation. In contrast, if the film nucleates as smaller domains in which a$_{MoO3}$ is parallel to either a$_{STO}$ or b$_{STO}$, the RHEED streaks along both [100] and [010] are expected to split into two components corresponding to [100]$_{MoO3}$ ∥ [100]$_{STO}$ and [100]$_{MoO3}$ ∥ [010]$_{STO}$).[27] Following reference 26, we refer to the latter structural configuration as an **iso-oriented** film. However, neither classification of RHEED patterns (large single domain or iso-oriented smaller domains) has been observed in films grown by MBE, sputtering, atomic layer deposition, or pulsed laser deposition (PLD). Rather, the initial layers were either porous, polycrystalline, of a different structure, or lacked a well-defined in-plane orientation.[23, 27, 29, 31-35]

In this work, we show that well-defined α-MoO$_3$(010) thin films with monolayer thickness (0.7 nm) can be reproducibly grown on STO substrates by MBE through a self-limiting process. The good lattice match along *a* axis ensures that iso-oriented α-MoO$_3$ films will align with STO along either [100] or [010] direction. The thickness and quality of the films were found to strongly depend on growth temperature. We have identified optimal growth conditions which lead to the nucleation of monolayer α-MoO$_3$. Further deposition following completion of the first monolayer does not lead to an increase in thickness as the thermal energy results in desorption of additional layers. These layers are bound only by weak van der Waals interaction, resulting in a self-limiting growth mechanism. Once the initial iso-oriented α-MoO$_3$ monolayer is formed, lowering the growth temperature by 50$^o$C leads to iso-oriented, high-quality, thicker α-MoO$_3$ films.

**Results and discussion**

RHEED was used to monitor the morphology and crystallinity during the film growth process. The starting patterns of STO(001) taken along [100] and [110] azimuthal directions are shown in Fig. 1b. The RHEED beam was blanked during the majority of deposition time and the patterns along [100] were taken periodically with minimal exposure to avoid beam-induced sample reduction.[36] As found previously, film structure and crystallinity proved to be critically dependent on growth temperature.[23, 27-29] We found that when the substrate temperature is less than 400$^o$C, the films remain amorphous as judged from RHEED patterns as shown in the Supporting information (Fig. S1). Deposition at 450$^o$C results in the nucleation of iso-oriented α-MoO$_3$ monolayers, although full crystallization into a well-ordered structure required 30 min of annealing. Nanoflakes (0.7 - 2.6 nm in height , 50 - 200 nm in width marked by blue arrows) and small islands (8 - 15 nm in height, 60 - 100 nm in height marked by white arrows) are observed to distribute randomly across the surface at higher temperatures (500°C) duo to secondary phase formation as judged from the RHEED patterns and atomic force microscopy (AFM) imaging (Fig. S1g and i). At 550°C, sharp but spotty RHEED patterns are observed which are correlated with isolated flat-top nanoislands (3 - 5 nm in height, 30 - 150 nm in width) as seen in AFM scans (Fig. S1j). XPS data (not shown) for all films shown in Fig. S1 indicate that the area-averaged film thickness is less than 1 nm even though the nominal thicknesses estimated from the deposition rate and time should be 39.6 nm at 450°C, 9.7 nm at 500°C, and 9.4 nm at 550°C. These discrepancies were observed earlier in MBE growth of $MoO_3$ and were ascribed to re-evaporation resulting from the high substrate temperature.[23] We will show that this self-limiting process is the key to achieving atomically flat, well-ordered α-$MoO_3$.

The film shown in Fig. 1b was grown at a substrate temperature of 450°C. The RHEED patterns recorded at 40 min (14.4 nm nominal thickness) were identical to those taken at the end of the film growth (110 min, ~40 nm nominal thickness) as shown in Fig. 1b. The actual film thickness is less than a nanometer as verified by both XPS and STM as shown in Fig. 2 and 3, and will be discussed later. The sharp streaks along both [100] and [110] are aligned parallel to those of the substrate, indicating the epitaxial relationship and a smooth film surface. The most striking difference from previous studies is the clear splitting of the streaks observed along [100] and [010] azimuthal directions as indicated by the red arrows. Using the STO substrate as an internal standard, the in-plane lattice parameters for the $MoO_3$ film can be estimated from the line profiles as shown in Fig. 1b. While one peak in the doublet matches that of STO (3.905 Å), the other equally intense peak corresponds to 3.69 Å, consistent with the smaller (*c*) in-plane lattice parameter of α-$MoO_3$(010). In addition, the RHEED patterns taken along [100] and [010] are identical, pointing to epitaxy with an iso-oriented alignment, i.e., $(010)_{MoO3} \parallel (001)_{STO}$, and $[100]_{MoO3} \parallel [100]_{STO}$ or $[010]_{STO}$.

High-resolution XPS core-level Mo 3d, O 1s, Ti 2p, and Sr 3d spectra for an ultra-thin film grown on STO is shown in Fig. 2. The Mo 3d spectrum for stoichiometric MoO$_3$ (Mo$^{6+}$) should be composed of a well-resolved Mo 3d$_{5/2}$ (232.5 eV) and 3d$_{3/2}$ doublet.[37-40] For our spectrum, accurate peak fitting requires a second doublet shifted 1.1 eV to lower binding energy, revealing the presence of a small amount of Mo$^{5+}$.[39, 40] The Mo$^{5+}$ peak area is ~9% of the total, suggesting a formula of MoO$_{2.95}$. The strong Ti 2p and Sr 3d signals detected from the substrate are clearly inconsistent with the nominal thickness of 39.6 nm expected based on the MoO$_3$ flux. In order to better estimate the thickness of the ultra-thin film, we measured survey spectra 70° off normal. These are displayed together with normal emission spectra for clean STO, the ultra-thin film, and a 40 nm α-MoO$_3$ film (confirmed by X-ray diffraction and measured by X-ray reflectivity, as will be discussed later) in Fig. 2e. Ti 2p and Mo 3p are shown within 390 – 480 eV range. While little change in Mo 3p intensity was observed between normal emission and 70° off normal, the intensity of Ti 2p was dramatically decreased when measured at the glancing angle, indicating good surface wetting. Using a previously developed method for measuring the thickness of oxide films proposed by Hill et al.,[41, 42] we estimate the film thickness to be ~0.8 – 1.1 nm, close to one monolayer of α-MoO$_3$ (0.70 nm).

Referring back to the Mo 3d spectrum, the slight reduction could be due to the charge transfer between the first α-MoO$_3$ monolayer and the STO surface. As shown in Fig. 1a, each α-MoO$_3$ layer is terminated by O atoms. In addition to the van der Waals bonding, charge transfer could occur between those O atoms and defect sites or dangling bonds on STO surfaces. Similar charge transfer has been observed between ultra-thin layers of MoO$_3$ and other support, such as Au and graphene.[38, 43] In comparison, the Mo 3d spectrum taken from a 40 nm α-MoO$_3$ film grown at 400°C (see inset of Fig. 4e, discussed below) can be perfectly fit by a single doublet, supporting the notion that sample reduction is due to an interfacial effect. We propose that charge transfer stabilizes the first monolayer. The subsequent layers immediately desorb due to weaker van der Walls bonding, resulting in a self-limited process. It should also be noted that prolonged annealing (30 min) during deposition is needed for the first layer to evolve, reveling the growth process is strong intertwined with thermodynamics and kinetics.

We have also directly imaged an ultra-thin α-MoO$_3$ film using STM (Omicron VT). The film was grown *in situ* on a Nb-doped STO(001) substrate. As shown in Fig. 3a, the ultra-thin film consists of atomically flat MoO$_3$ nanoflakes (films) which are oriented along <100> directions. The height of the film measured from the line profile shown in Fig. 3b is ~0.7 nm, confirming the ultra-thin α-MoO$_3$ film is indeed one monolayer thick. A single-step defect terrace structure on the STO substrate is also displayed as a reference (blue dashed line), and has a height of 0.2 nm.

In order to prevent re-evaporation and grow the film to a higher thickness, we lowered the substrate temperature to 400°C after the initial ultra-thin layer was formed at 450°C. The RHEED patterns taken at 10 and 40 nm nominal thicknesses are shown in Figs. 4. These patterns are clearly different from those observed for the ultra-thin layers, and yet still reveal an epitaxial orientation to STO. Spotty patterns observed for the 10 nm film are most likely a result of 3D island nucleation on the iso-oriented α-MoO$_3$ domains. At 40 nm, the pattern becomes streakier, indicating a smoother surface. Moreover, the streak splitting re-appears in the RHEED pattern taken along [100] azimuthal direction, indicating that the thicker α-MoO$_3$ islands are still iso-oriented. The film is fully stoichiometric, consisting of only $Mo^{6+}$ as judged from the Mo 3d spectrum shown in the inset of Fig. 4e. The out-of-plane XRD θ-2θ scan for the 40 nm film (Fig. 4e) contains a single set of (*0l0*) α-MoO$_3$ film peaks along with the sharper STO substrate peaks, revealing single-phase epitaxy. The values of the full width at half maximum (FWHM) in the high-resolution X-ray rocking curves (Fig. 4f) are 0.01° for both the film and the STO substrate, indicating that the α-MoO$_3$ film is highly crystalline, with its overall structural quality being limited by the substrate. In comparison, α-MoO$_3$ films grown by atomic layer deposition (ALD) on Si(111) displayed much wider FWHM (~4.5°),[29] which was ascribed to tilt in the fibrous or columnar texture of the sample. We argue that the initial high-quality, iso-oriented ultra-thin layer grown on STO serve as a template for the successive growth, ensuring iso-oriented epitaxy and high crystallinity.

**Experimental**

The MoO$_3$ thin films were grown on double-side polished STO(001) in a custom MBE system described elsewhere.[44] High-purity MoO$_3$ powders (Sigma-Aldrich, >99.99%) were evaporated from an effusion cell at a film growth rate of ~0.1 Å/s as calibrated by a quartz crystal microbalance (QCM). The STO substrates (10×10×0.5 mm, MTI Corporation) were rinsed in DI H$_2$O for 30 seconds and annealed in a tube furnace in air at 1000°C for 8 hrs. They were subsequently cleaned by heating in the MBE chamber at 600°C for 20 min in an oxygen partial pressure of 6.0 × 10$^{-6}$ Torr prior to film growth. The substrate temperature during growth was varied from 400 to 550°C, and 450°C was determined to be the optimal condition for the growth of iso-oriented α-MoO$_3$ films with monolayer thickness. An activated oxygen plasma beam (with O$_2$ partial pressure in the chamber set at ~3×10$^{-6}$ Torr) was incident on the sample during deposition to prevent sample reduction. *In situ* RHEED was used to monitor the overall morphology and surface structure. After deposition, the substrate temperature was reduced at a rate of 10 °C min$^{-1}$ under the same oxygen environment. *In situ* high-resolution XPS using monochromatic Al K$_{\alpha 1}$ x-rays ($hv$ = 1486.6 eV) was carried out at an electron take-off angle of 0º relative to the surface normal unless otherwise noted with a VG/Scienta SES 3000 electron energy analyzer in an appended chamber. The total energy resolution was 0.5 eV. MoO$_3$ films were also grown *in situ* in an Omicron VT STM system for high resolution imaging. Nb-doped STO (0.05 weight %, Crystec Corporation) was used to achieve stable tunneling current. STM images were acquired with a commercial Omicron W tip. The high-resolution XRD scans were recorded using a Philips X'Pert Materials Research Diffractometer equipped with a fixed Cu anode operating at 45 kV and 40 mA, a hybrid monochromator consisting of four-bounce double crystal Ge(220), and a Cu X-ray mirror. A Dimension Icon atomic force microscope (AFM, Bruker, USA) was employed in contact mode to acquire *ex situ* AFM images shown in Fig. S1. Au coated silicon nitride probes (Bruker) were used and the scanning rate was 0.25 Hz.

**Conclusions**

In summary, we demonstrate the growth of monolayer-thick iso-oriented α-MoO$_3$ films by molecular beam epitaxy through a self-limiting process. While the charge transfer between ultra-thin MoO$_3$ film and STO substrate stabilizes the interfacial layer, the successive growth is inhibited by re-evaporation at a growth temperature of 450ºC. The charge transfer observed between ultra-thin α-MoO$_3$ and STO may significantly alter the work function, band structure, and band alignment of MoO$_3$, creating additional opportunities to tune the properties of these ultra-thin films.[38, 43] Lowering the growth temperature after the growth of the interfacial layer leads to thicker phase-pure, iso-oriented α-MoO$_3$ with excellent crystallinity. The ability to synthesize these high-quality, phase-pure, epitaxial films under UHV conditions will allow *in-situ* characterization and enable more fundamental research to be carried out on their surfaces.

**Acknowledgements**

YD acknowledges support by EMSL's Intramural Research and Capability Development Program. A portion of the work was supported by the US DOE Office of Basic Energy Sciences, Division of Materials Science and Engineering under Award 10122. ZD, IL, and RM acknowledge the support by the US DOE Office of Basic Energy Sciences, Division of Chemical Sciences, Geosciences & Biosciences. GL acknowledges support by Henan University, China. EWP and JZ are supported by National Science Foundation (Award Number: CHE1151846). The work was performed at the W. R. Wiley Environmental Molecular Sciences Laboratory, a DOE User Facility sponsored by the Office of Biological and Environmental Research.

**Notes and references**

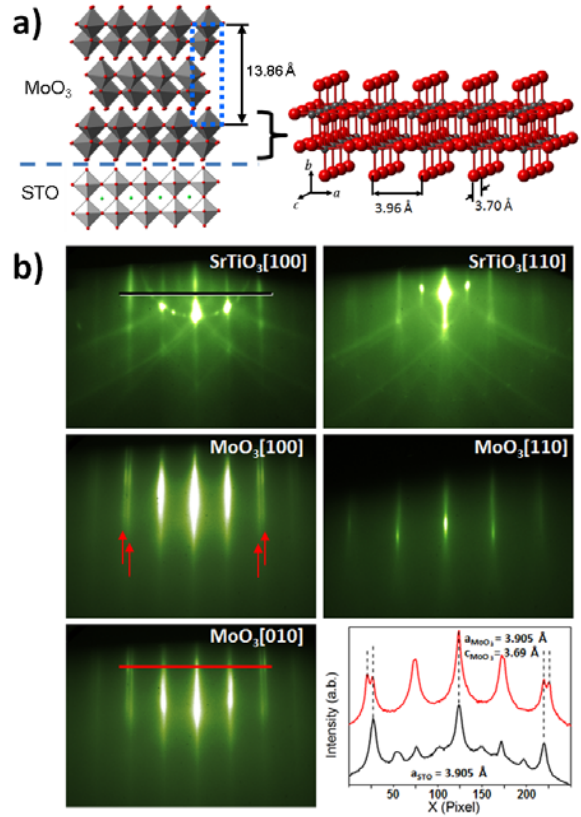

Figure 1. (a) Structural models of α-MoO$_3$ and SrTiO$_3$, with MoO$_6$ and TiO$_6$ octahedra shown. Red circles represent oxygen atoms and green circles represent Sr atoms. Crystalline α-MoO$_3$ are composed of bilayer sheets stacked in the [010] direction. (b) RHEED patterns and line profiles taken along different azimuthal directions for the SrTiO$_3$(001) substrate and an α-MoO$_3$ thin film.

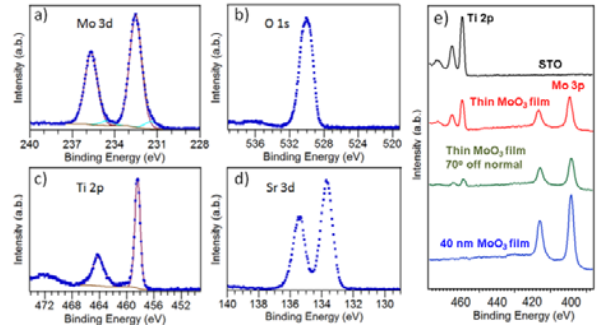

Figure 2. High-resolution XPS core-level spectra obtained at normal emission: Mo 3d (a), O 1s (b), Ti 2p (c) and Sr 3d (d) spectrum for an ultra-thin $MoO_3$ film grown on STO at 450°C. (e) XPS survey spectra including the Ti 2p and Mo 3p peaks for an STO substrate, an ultra-thin $MoO_3$ film grown at 450°C, and a thicker $MoO_3$ film grown at 400°C.

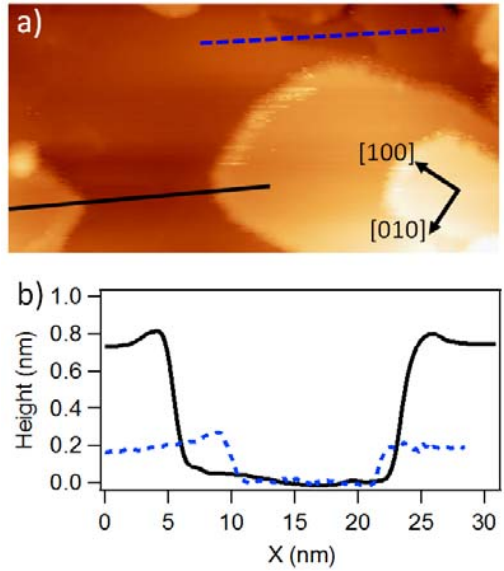

Figure 3. AFM image (70nmx35nm) of an α-MoO$_3$(010) film grown on STO (a) and the profiles (b) of the lines marked in (a).

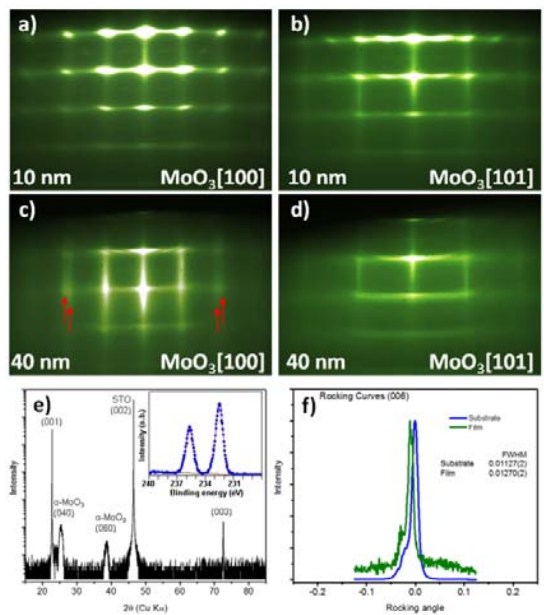

Figure 4. (a-d) RHEED patterns taken along [100] and [101] azimuthal directions of a MoO$_3$ film grown to 10 and 40 nm thicknesses, respectively. XRD θ-2θ scan (e) and rocking curve (f) for the 40 nm film. The inset of Fig. 4e displays the XPS core-level Mo 3d spectrum for the 40 nm film.

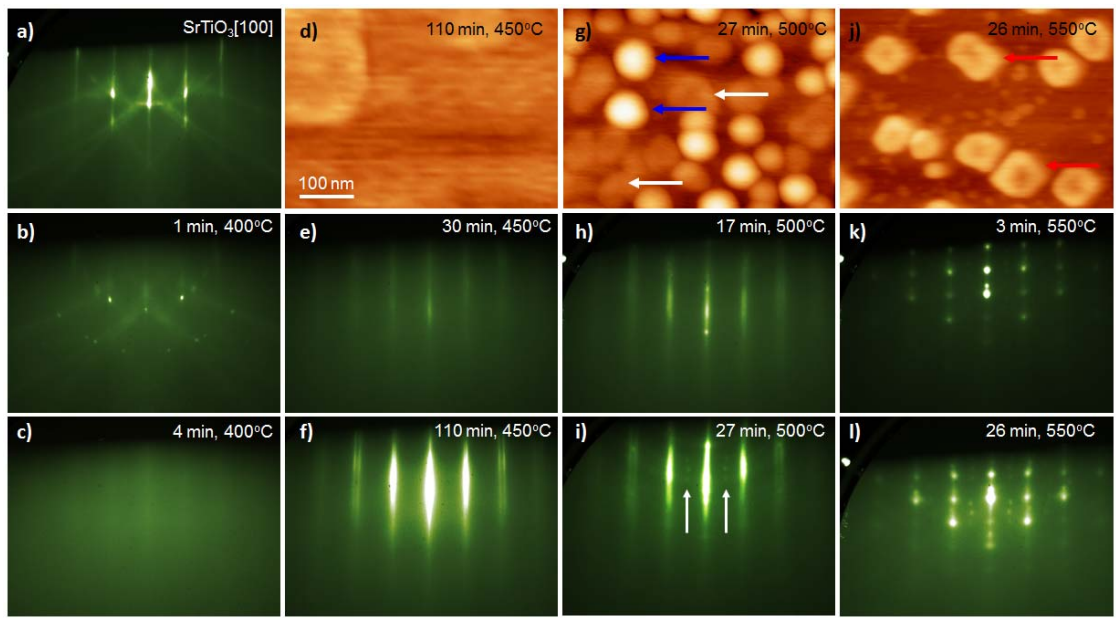

Figure s1: RHEED patterns taken along [100] azmuthal direction and AFM images (d, g, and j) of MoO$_3$ films grown on SrTiO$_3$(001). Total deposition time and substrate tempearture for each film are marked.